# Role of E-F Coupling on Low-latitude Ingression of Mid-latitude MSTIDs over the Indian Subcontinent

Jeevan Upadhyaya[1], Dipjyoti Patgiri[1,2], Mehta Harsh Jayeshkumar[1], Rahul Rathi[3], Virendra Yadav[4], Satarupa Chakrabarti[5], Arun K. Upadhayaya[6], Sumanta Sarkhel[*1,7]

[*]Sumanta Sarkhel, Department of Physics, Indian Institute of Technology Roorkee, Roorkee – 247667, Uttarakhand, India (sarkhel@ph.iitr.ac.in)

[1]Department of Physics, Indian Institute of Technology Roorkee, Roorkee – 247667, Uttarakhand, India.

[2]Heliophysics Division, NASA Goddard Space Flight Center, 8800 Greenbelt Rd, Greenbelt, MD 20771, USA.

[3]School of Physics and Astronomy, Lancaster University, Lancaster, United Kingdom.

[4]Aryabhatta Research Institute of Observational Sciences, Nainital – 263001, Uttarakhand, India.

[5]Department of Computer Science & Engineering, Motilal Nehru National Institute of Technology Allahabad, Prayagraj – 211004, Uttar Pradesh, India.

[6]Environmental Sciences and Biomedical Metrology Division, CSIR National Physical Laboratory, New Delhi, India.

[7]Centre for Space Science and Technology, Indian Institute of Technology Roorkee, Roorkee –247667, Uttarakhand, India.

## Abstract

This study provides the first observational evidence that E-F region electrodynamic coupling regulates the southernmost latitudinal extent of nighttime mid-latitude medium-scale traveling ionospheric disturbances (MSTIDs) over the Indian subcontinent. We have analyzed multiple MSTID events using total electron content (TEC) measurements from a dense network of dual-frequency GNSS receivers across India, GNSS radio occultation (RO) profiles, and ionosonde observations from New Delhi during May-July 2025, a high solar activity period. The lowest geographic latitude, at which MSTIDs' wavefronts reached, varied between ~17°N and ~25°N. Their low-latitude extension showed no significant correlation with background TEC, but was positively correlated with the geomagnetically mapped southernmost latitude of sporadic-E (*Es*) layers identified from GNSS-RO measurements. These events occurred during high solar activity, when F-region MSTID growth is generally suppressed. Therefore, our results suggest that electrodynamic coupling with the underlying *Es*-layers is essential for sustaining MSTIDs and enabling their low-latitude penetration during solar maxima.

**Keywords:** MSTIDs, SOI-CORS GNSS Network, Sporadic-E (*Es*), E-F Coupling, Low-latitude Ingression of MSTIDs.

## Key Points:

1. Nighttime MSTIDs over the Indian Subcontinent show varying low-latitude ingression that does not correlate with background VTEC.
2. Our observations suggest that the sporadic-E layer mediated E-F coupling influences the southernmost latitude extent of nighttime MSTIDs.
3. Suppressed Perkins instability during high solar activity may explain why *Es* activity governs low-latitude MSTID ingression.

**Plain Language Summary**

Nighttime Medium-scale traveling ionospheric disturbances (MSTIDs) are propagating wave-like fluctuations in electron density in the Earth's F-region ionosphere. These disturbances can affect radio wave propagation, thereby affecting GNSS systems. During nighttime, MSTIDs manifest as electrified structures associated with polarization electric fields that typically drive their equatorward propagation. While MSTIDs are primarily considered mid-latitude phenomena, they can occasionally reach lower latitudes. However, the electrodynamic factors that govern their low-latitude ingression limits remain unclear. In this study, we utilized an unprecedented dense network of more than 900 GNSS receivers across India to track MSTIDs' propagation over the Indian subcontinent during the summer months (May-July) of 2025 (solar maximum period). It is important to note that the solar maxima period generally create less favorable condition for MSTID development and their propagation to lower latitudes. However, our observations reveal that low-latitude MSTIDs' propagation is actively sustained by another plasma irregularity in the E-region of the ionosphere known as the Sporadic-E (*Es*) layer. This is because the two altitude regions are electrodynamically coupled through the Earth's geomagnetic field lines and provide essential positive feedback to overlying MSTIDs. This demonstrates that vertical electrodynamic coupling between different ionospheric layers is critical for sustaining MSTIDs at lower latitudes.

## 1. Introduction

Medium scale traveling ionospheric disturbances (MSTIDs) are propagating bands of high and low plasma density in the Earth's ionosphere and are characterized by horizontal wavelengths of a few hundred kilometers. Over mid-latitudes, MSTIDs have prominently been observed to propagate towards equator during geomagnetically quiet times (Afraimovich et al., 2013; Astafyeva, 2019; Ding et al., 2011; Duly et al., 2013; Fišer et al., 2017; Hernández-Pajares et al., 2006). During nighttime, MSTIDs are primarily driven by an electrodynamic instability mechanism, commonly known as Perkins instability (Hamza, 1999; Perkins, 1973). The linear growth rate of this instability is inversely proportional to the ion-neutral collision frequency and increases (decreases) during low (high) solar activity (Otsuka et al., 2021). These electrodynamically generated MSTIDs exhibit northwest-southeast aligned phase fronts consisting of polarization electric fields, which propagate southwestward in the northern hemisphere (Ding et al., 2011; Duly et al., 2013; Kotake et al., 2006; Otsuka et al., 2007; Patgiri et al., 2024c; Shiokawa et al., 2003).

Although, Perkins instability explains the preferential alignment and associated polarization electric fields of nighttime MSTIDs, its inherently low linear growth rate ($\sim 10^{-4}$ $s^{-1}$) remains a major limitation in explaining the development of MSTIDs (Kelley & Makela, 2001; Narayanan et al., 2018; Otsuka et al., 2007). Theoretical and observational studies have demonstrated that E-F region coupling significantly enhances this low linear growth rate of Perkins instability by providing positive feedback (Cosgrove et al., 2004; Narayanan et al., 2018; Otsuka et al., 2007; Tsunoda, 2006; Tsunoda & Cosgrove, 2001; Yokoyama et al., 2009). In the ionospheric E-region, the sporadic-E ($E_S$) layer is a common irregularity feature which is generated due to the wind shear mechanism and can have polarization electric fields associated with it (Haldoupis, 2011; Yamazaki et al., 2022). These polarization electric fields can map along geomagnetic field lines into the F-region, thereby reinforcing the Perkins instability and promoting the growth of MSTDs (Narayanan et al., 2018; Otsuka et al., 2007; Patgiri et al., 2024c).

At mid-latitudes, MSTIDs generated by Perkins instability preferentially propagate equatorward. Although, only a few cases of their propagation into low-latitude regions have been reported (Makela et al., 2010; Narayanan et al., 2014; Shiokawa et al., 2002; Sivakandan et al., 2019). Shiokawa et al. (2002) suggested that the usually higher ion drag in the equatorial ionization anomaly (EIA) region can influence the low-latitude propagation of MSTIDs. Makela et al. (2010) reported that solar minima conditions favor low-latitude ingression of MSTIDs due to reduced background electron density and lower ion drag. Narayanan et al. (2014) and Sivakandan et al. (2019) suggested that a weak EIA, absence of midnight brightness wave (MBW), and solar minimum conditions can play a role in the propagation of MSTIDs to low-latitudes. All of these studies have reported the role of various ionospheric/thermospheric background conditions on the deep equatorward propagation of MSTIDs. However, potential role of $E_S$-layers and the associated E-F region coupling in facilitating the propagation of MSTIDs to low-latitudes has not been explicitly explored and reported in the literature.

The present study uses GNSS-TEC maps generated from a dense receiver network over India to identify the nighttime MSTIDs. These observations are analyzed alongside $E_S$-layer detections from SNR profiles of multiple GNSS radio occultation (RO) missions, complemented by ionosonde measurements, to investigate how E-F region coupling influences MSTID propagation. This study will provide a basis for future investigations into the equatorward propagation of mid-latitude MSTIDs and their potential influence on low-latitude and equatorial ionospheric dynamics.

## 2. Instruments and Methodology

We utilized a dense network of Continuously Operating Reference Stations (SOI-CORS) of GNSS receivers maintained by the Survey of India (Singh & Kumar, 2019). By the summer of 2025, this network comprised over 900 active receivers (Figure 1a). This spatial coverage enables high-resolution mapping of ionospheric perturbations over the Indian subcontinent. Dual-frequency carrier phase and pseudo-range measurements were sampled at a 30-second interval, which were processed to extract Slant and Vertical Total Electron Content (STEC and VTEC) following the methodology of Seemala (2023), assuming an Ionospheric Pierce Point (IPP) altitude of 350 km. An elevation cutoff of 35° was applied to mitigate errors associated with the thin-shell approximation. Following established methodology (Otsuka et al., 2013; Tsugawa et al., 2007), we have isolated wave-like signatures of MSTIDs. We computed detrended VTEC (DVTEC) by subtracting the 1-hour running average (±30 min) from the background VTEC. The calculated DVTEC values were then spatially gridded at a 0.25°×0.25° resolution. Finally, for compensating the data gaps, the 30s VTEC and DVTEC data were aggregated into a 3-minute average (±90s) and interpolated via a 2D moving mean across a 3×3 window. This processing reduced the average grid-wise standard deviation of collocated measurements (different PRNs and receivers) from around 1 TECU (raw VTEC) to 0.1 TECU (DVTEC), well below typical >0.2 TECU MSTID amplitudes (Kotake et al., 2007; Otsuka et al., 2013). Our analysis focuses specifically on nighttime MSTIDs during the high solar activity period of May-July 2025. The events are characterized using the mean $\Delta_{10.7P}$ proxy calculated by averaging the daily F10.7 flux with its 81-day historical mean (Lei et al., 2005; Patgiri et al., 2024a; Yadav et al., 2017). In the present study, we have analyzed 21 different MSTID events in geomagnetically quiet conditions and presented two nights with contrasting ingression: 20 May 2025 ($K_P$ ~3) and 27 May 2025 ($K_P$ ~2). The remaining 19 events are presented in the supplementary material (Figures S1-S19). For all the events, the analysis covers the period from 13:00 to 24:00 UTC, corresponding to 18:30-05:30 IST (IST = UTC + 5.5 h).

The presence of *Es*-layers was detected using GPS L1 signal-to-noise ratio (SNR) profiles obtained from the RO measurements of COSMIC-2, SPIRE, MetOp-B and C, following the techniques described by Patgiri et al. (2024b), Tsai et al. (2018) and Yamazaki et al. (2022). RO tangent points were filtered to retain only nighttime profiles within the Indian region (68°-97°E; 7°-37°N). The RO-SNR profiles were normalized against a 20 km running mean background, and we computed their 1 km running standard deviation to derive the amplitude scintillation index ($S_4$) (Patgiri et al., 2024b). An *Es*-layer was confirmed when $S_4 \geq 0.2$ strictly within 90 to 120 km altitude (Tsai et al., 2018). To establish the spatial relationship governing E-F region coupling, the E-region RO tangent point locations were mapped to the 350 km IPP altitude along geomagnetic field lines using the Altitude-Adjusted Corrected Geomagnetic version 2 (AACGM-v2) model (Shepherd, 2014). Since the RO profiles only provide intermittent observations in space and time, for continuous ground-based verification of *Es*-layer occurrence, we also utilized ionograms from New Delhi (28.6°N, 77.2°E; Upadhayaya & Mahajan, 2013) whose location is marked by a black asterisk in Figure 1a. The corresponding ionograms were visually inspected to confirm the presence or absence of *Es*-layers, identified by a characteristic blanketing type layer near 100 km virtual height. The following section presents the results derived from the analyses of these datasets.

## 3. Results

### 3.1 Variability in the equatorward propagation limits of MSTID wavefronts

Figures 1b-1g and 1h-1m present the VTEC and DVTEC maps, respectively, for two geomagnetically quiet nights: 20 May 2025 ($K_P$~3) and 27 May 2025 ($K_P$~2). The MSTIDs observed in the DVTEC maps on these two nights propagated to different equatorward latitudes (Figure 1h-1m). On 20 May 2025, the MSTID wavefronts propagated into the low latitudes, reaching as far south as ~17°N geographic latitude, extending across nearly the entire Indian mainland (Figures 1h-1j). However, on 27 May 2025, the MSTID remained confined to ~25°N geographic latitude (Figures 1k-1m). The complete temporal and spatial evolution of the observed MSTIDs, together with the background VTEC for both nights, is provided as a supplementary movie (Movie S1). On both nights, MSTID activities were observed after 16:00 UT (21:30 IST), when the Equatorial Ionization Anomaly (EIA) generally weakens or decays over the Indian sector (Kumar et al., 2021, 2022; Sunda & Vyas, 2013). The corresponding VTEC maps also show that the northern EIA crest had already decayed with no distinct crest visible during this period (Figures 1b-1g). Although the background VTEC was slightly higher on 27 May 2025 as compared to 20 May 2025, a detailed analysis of all 21 events using VTEC keograms shows no consistent trend between background VTEC and equatorward extent of MSTIDs (Figure 2). Here, the equatorward extent of a MSTID event is defined as the southernmost geographic latitude reached by its wavefronts within the observational field-of-view during that night. The scatter plot in Figure 2 also shows no clear correlation ($r$ ~0.1; $p$ ~0.66) between the MSTIDs' equatorward extent and the mean VTEC within a 5°×5° window around that location and at the corresponding time.

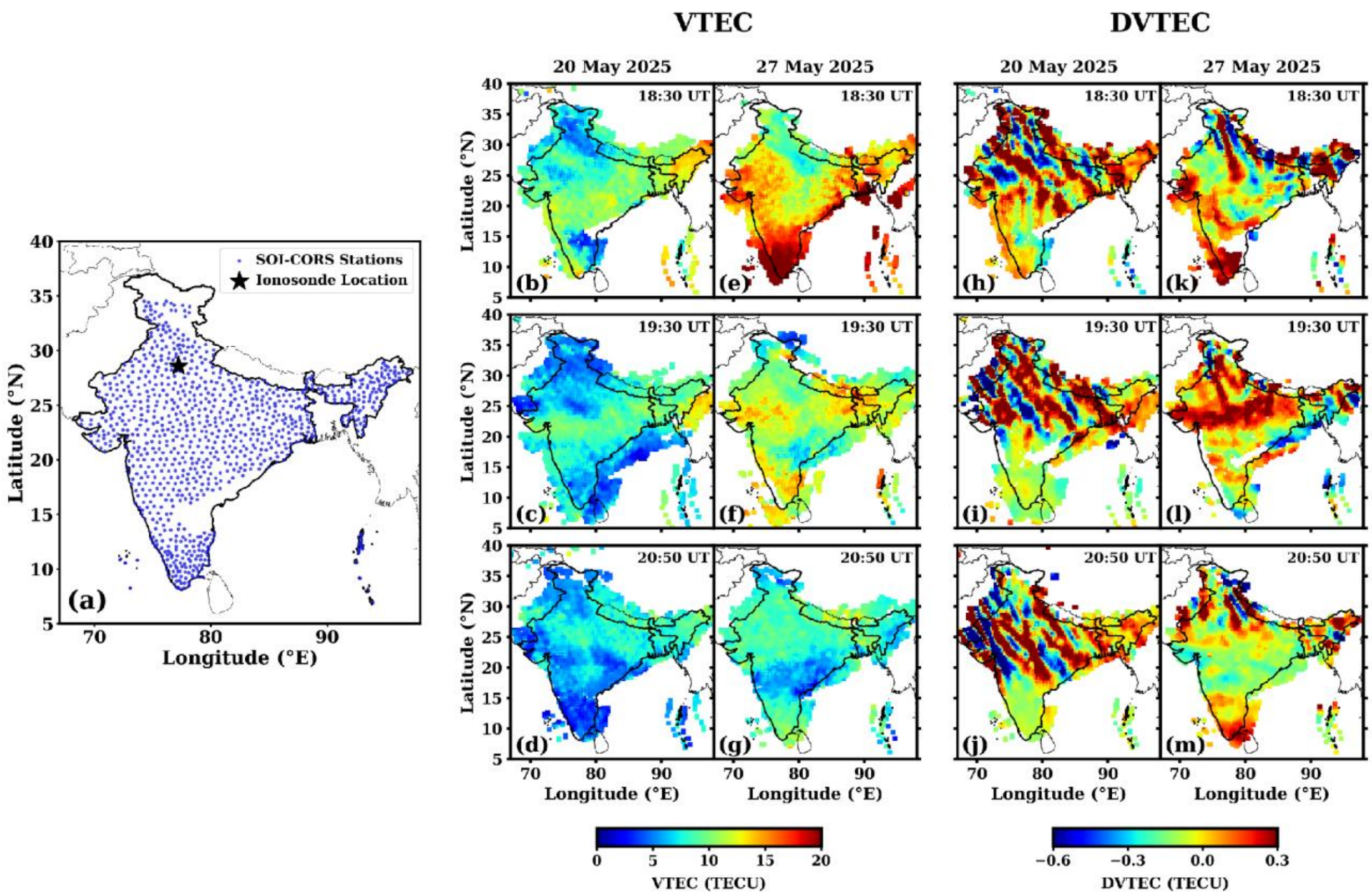

**Figure 1.** (a) Locations of the SOI-CORS GNSS receivers and the New Delhi ionosonde. (b-g) VTEC and (h-m) DVTEC maps for 20 and 27 May 2025.

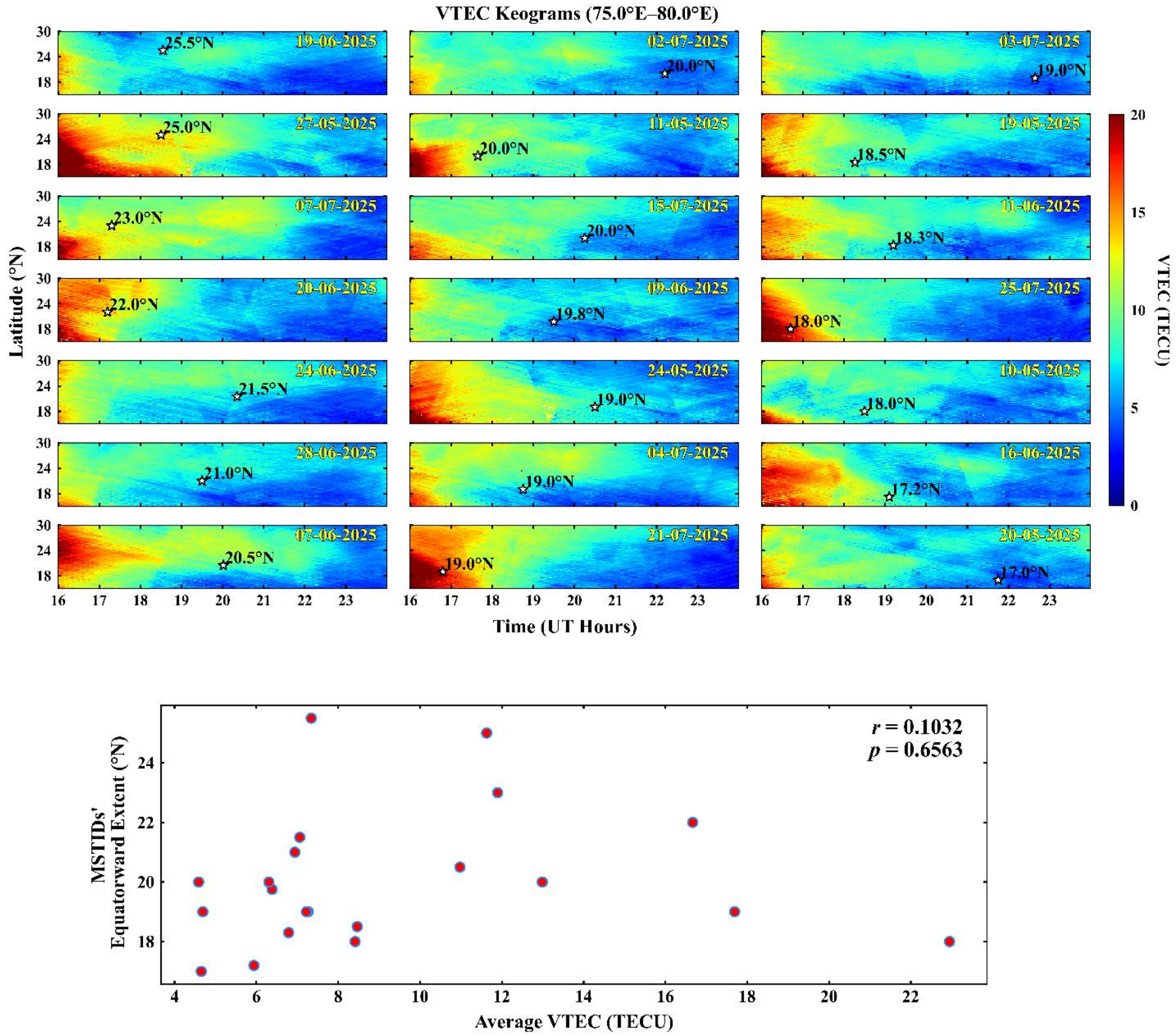

**Figure 2.** Upper panels: VTEC keograms for 75-80°E longitude during 21 nighttime MSTID events observed in May-July 2025. The asterisks indicate the southernmost latitude extent of MSTID wavefronts. Bottom panel: Scatter plot of the southernmost latitude extent versus mean background VTEC within a 5°×5° window (±2.5° in latitude and longitude) centered on the southernmost wavefront location at the corresponding time.

## 3.2 Detection of sporadic-E layers

This study utilizes RO-SNR profiles and New Delhi ionosonde observations to identify *Es*-layers for all the 21 events. Figures 3 and 4 show the observations on 20 and 27 May 2025, respectively, while the remaining 19 events are provided in supplementary Figures S1-S19. The top panels of Figures 3 and 4 present the DVTEC maps overlaid with the detected RO tangent points in the E-region (90-120 km; white-filled circle with magenta boundary), their footprints mapped along the geomagnetic field lines to the 350 km IPP altitude (white-filled circle with black boundary), and the New Delhi ionosonde's location (black asterisk). The middle panels of Figures 3 and 4 display the corresponding normalized SNR and $S_4$ index profiles while the bottom panels show the ionograms during the MSTID events.

The observations demonstrate that the southernmost mapped locations of the detected *Es*-layers ($S_4 \geq 0.2$) differed between the two nights and matched the equatorward extent of the MSTID wavefronts. On 20 May 2025, *Es*-layers were observed over the northern latitudes (Figures 3i-3j) as well as along the equatorward propagation path of the MSTID (Figures 3a-3h). The New Delhi ionograms also showed a blanketing type *Es*-layer near 100 km during this event (Figures 3m-3p). Further south, no *Es*-layer were detected in the available RO profiles whose mapped footprint lay south of the southernmost latitude (~17°N) reached by the MSTID wavefronts (Figures 3k-3l). This suggest a spatial correspondence between the mapped distribution of detected *Es* layers and the equatorward extent of MSTIDs. On 27 May 2025, MSTID wavefronts were confined to the northern region (>25°N), where *Es* activity was also observed (Figures 4c-4d, 4g-4h). However, unlike 20 May 2025, no *Es* activity was detected along the expected equatorward propagation path of the MSTIDs (Figures 4a-4b, 4e-4f, 4i-4l). In addition, New Delhi ionosonde also didn't detect any *Es*-layer (Figures 4m-4p). As evident from the DVTEC on 27 May 2025, MSTID wavefronts didn't extend into the lower latitudes. A similar pattern is observed across the remaining 19 events (Supplementary Figures S1-S19), indicating that the equatorward extent of the MSTIDs closely matches the southernmost latitude of the detected *Es*-layers mapped to an altitude of 350 km along the geomagnetic field lines.

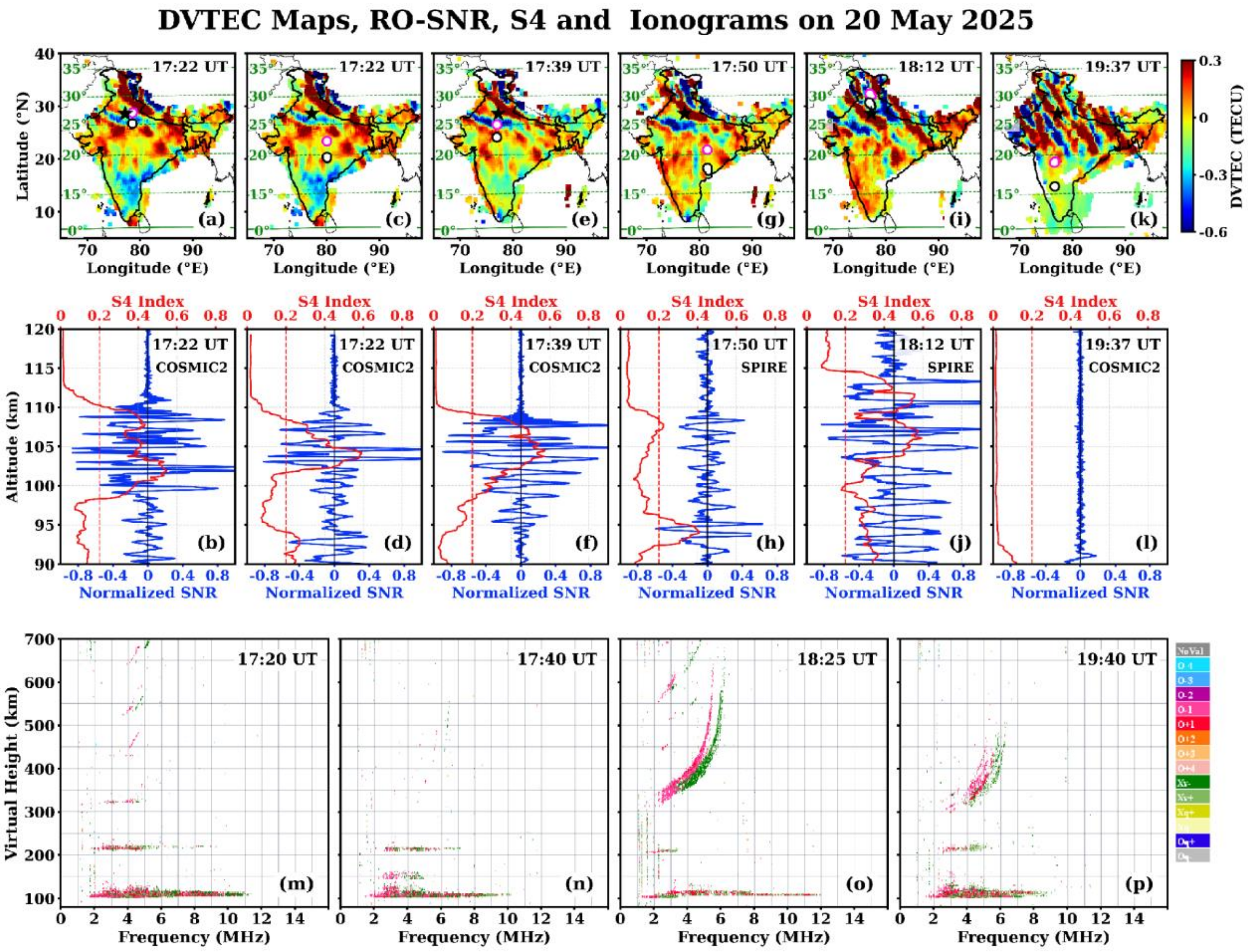

**Figure 3.** (a, c, e, g, i, & k) DVTEC maps for 20 May 2025 overlaid with E-region RO-SNR tangent points (white-filled circles with magenta boundaries), their mapped footprints at the 350 km IPP altitude (white-filled circles with black boundaries), and the New Delhi ionosonde's location (black asterisk). The blue and red curves in (b, d, f, h, j, & l) denote the corresponding profiles of normalized SNR and $S_4$, respectively. (m-p) The ionograms over New Delhi indicate the presence of *Es*-layers near 100 km virtual height.

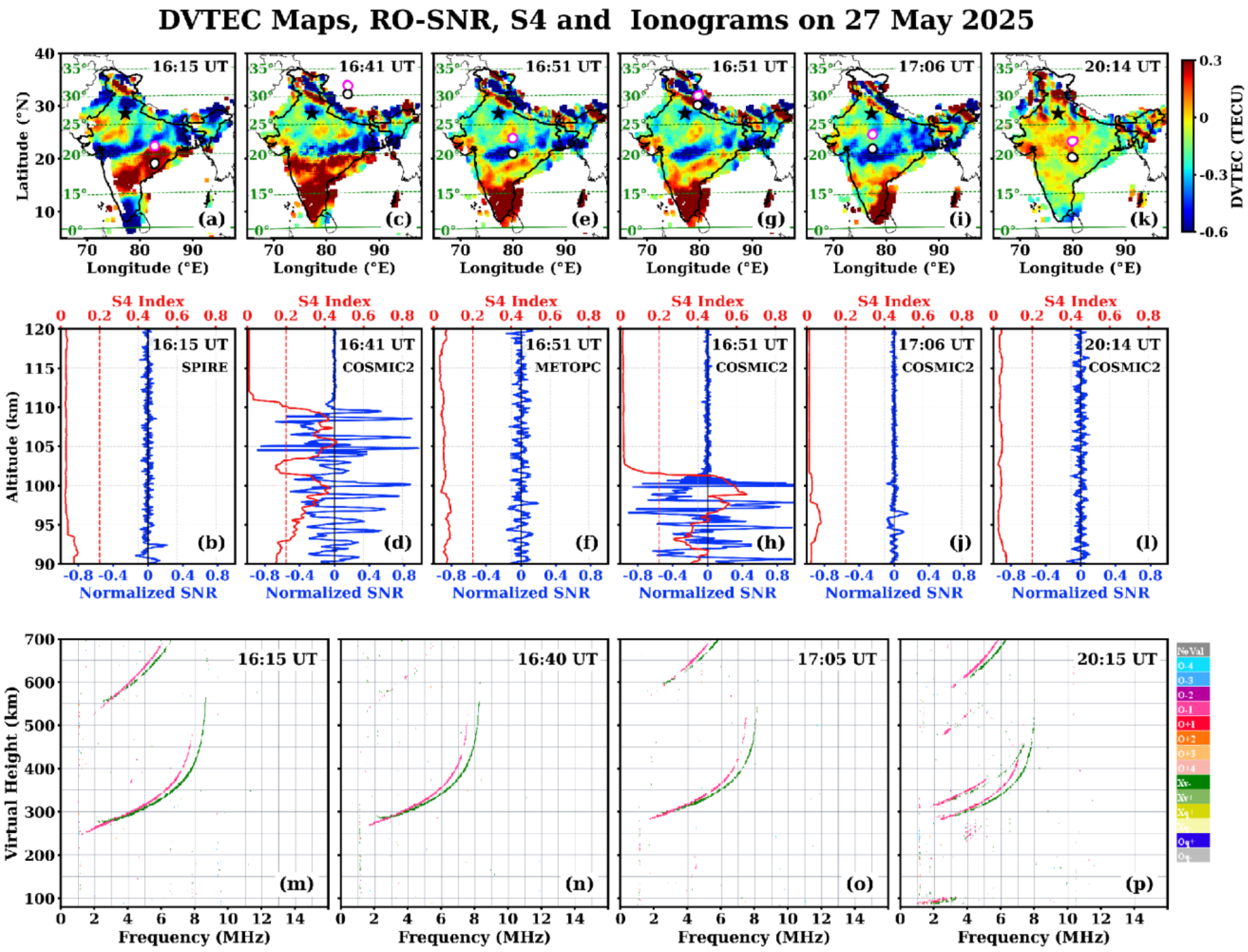

**Figure 4.** DVTEC maps, RO-SNR profiles and ionograms over New Delhi for the 27 May 2025 event. The format is identical to Figure 3. Note that Sporadic-E (*Es*) activity is absent ($S_4 < 0.2$) at the lower latitudes and no *Es* observed in the New Delhi ionograms.

## 4. Discussion

In the present study, TEC maps derived from a dense network of GNSS receivers across the Indian sector were used to investigate the equatorward extent of nighttime MSTIDs. We analyzed 21 MSTID events during May-July 2025 which exhibited northwest-southeast orientation and southwestward propagation. Their preferential alignment and propagation direction were consistent with the characteristic signatures of MSTIDs generated through the Perkins instability (Perkins, 1973; Hamza, 1999). However, the southernmost latitude reached by the MSTID wavefronts varied across these events from ~17°N to ~25°N geographic latitude.

Previous studies have suggested that enhanced electron density near the EIA crest is a primary factor inhibiting the equatorward propagation of MSTIDs (Shiokawa et al., 2002; Sivakandan et al., 2019). It is closely linked to growth rate of Perkins instability, that is directly proportional to effective electric field at the F-region ($\boldsymbol{E}^* = \boldsymbol{E} + \boldsymbol{U} \times \boldsymbol{B}$) and inversely proportional to the density weighted field-line-integrated ion-neutral collision frequency ($\langle \nu_{in} \rangle$) (Hamza, 1999; Otsuka et al., 2007, 2021a; Perkins, 1973). Here, $\boldsymbol{E}$ and $\boldsymbol{U}$ are the F-region dynamo electric field and thermospheric neutral wind, respectively. Enhanced F-region electron density near the EIA crest increases the field-line-integrated Pedersen conductivity, which shorts out the mapped polarization electric fields necessary to sustain nighttime MSTIDs (Kotake et al., 2007; Otsuka, 2021; Shiokawa et al., 2002; Sivakandan et al., 2019; Tsugawa et al., 2003, 2007). Enhanced plasma density also increases $\langle \nu_{in} \rangle$ and ion drag which further reduces the Perkins growth rate and damp atmospheric gravity waves that can seed the instability (Otsuka, 2021; Shiokawa et al., 2002; Tsugawa et al., 2007). However, our observations show that after the nighttime decay of the EIA crest, several MSTIDs remained confined to higher latitudes despite low background VTEC, while other events propagated into the lower latitudes under comparable or even higher background VTEC conditions. Furthermore, the VTEC keograms for all 21 events and the comparison between background VTEC and the lowest MSTID latitude show no correlation (Figure 2). This indicates that background F-region plasma density alone does not explain the equatorward extent of these MSTIDs. Instead, it corresponds closely to the spatial distribution of *Es*-layers observed in both the RO-SNR profiles and the New Delhi ionograms across the 21 events (Figures 3, 4, and S1-S19).

It is noteworthy that previous studies have reported a correlation between the presence of *Es*-layers, their associated inhomogeneities, and the occurrence of nighttime MSTIDs (Lee et al., 2021; Narayanan et al., 2018; Otsuka et al., 2008). The inhomogeneities in *Es* layers provide positive feedback to Perkins instability via E-F electrodynamic coupling and help in the generation of MSTIDs (Cosgrove & Tsunoda, 2004; Patgiri et al., 2024b; Tsunoda et al., 2006; Yokoyama et al., 2009). Additionally, the presence of *Es* activity allows MSTIDs to generate even when F-region background conditions are unfavorable for MSTID formation (Fu et al., 2026). Apart from contributing to the generation and sustaining the growth of MSTIDs, *Es*-layers can influence their propagation through E-F electrodynamic coupling. Numerical simulations by Yokoyama et al.

(2009) showed that rotational shear in meridional wind in the E region can produce southwestward phase propagation of coupled *Es* and F-region structures. On the other hand, observational study by Matsushima et al. (2022) found that *Es* structures and F-region MSTIDs generally propagated southwestward with similar speeds and directions, consistent with coupled propagation.

To illustrate E-F electrodynamic coupling mechanism, Figures 5a and 5b schematically show E-F coupling in two and three dimensions, respectively. The black strip-like layers represent the irregular *Es*-layers while the alternating blue and maroon fronts represent the MSTID wavefronts over the Indian subcontinent. Electric fields associated with irregular *Es*-layers map to the F region along geomagnetic field lines (***B***). The white-filled magenta and black circular outlines in Figures 5a & 5b denote the discrete E-region tangent points of RO-SNR measurements and their mapped F-region footprints, respectively. Since the geomagnetic field act as equipotential lines, these E-region electric fields associated with the irregularities in the *Es*-layers map to the F-region with negligible attenuation (Cosgrove & Tsunoda, 2004; Spreiter & Briggs, 1961). For 17 out of the 21 events, the RO observations were available when the MSTIDs reached their southernmost extent. The latitude of this extent was positively correlated ($r$ ~0.6; $p$ ~0.01) with the latitude of the southernmost observed *Es*-layer's footprint at 350 km (Figure 5c). However, the apparent not-so-strong correlation is likely due to the simultaneous RO profiles being unavailable (in some cases) close to the southernmost extent of MSTID fronts. The overall trend nevertheless suggests that MSTIDs extend to lower latitudes when active *Es*-layers are also observed at lower latitudes. In contrast, MSTIDs remain confined to higher latitudes when low-latitude *Es* activity is absent.

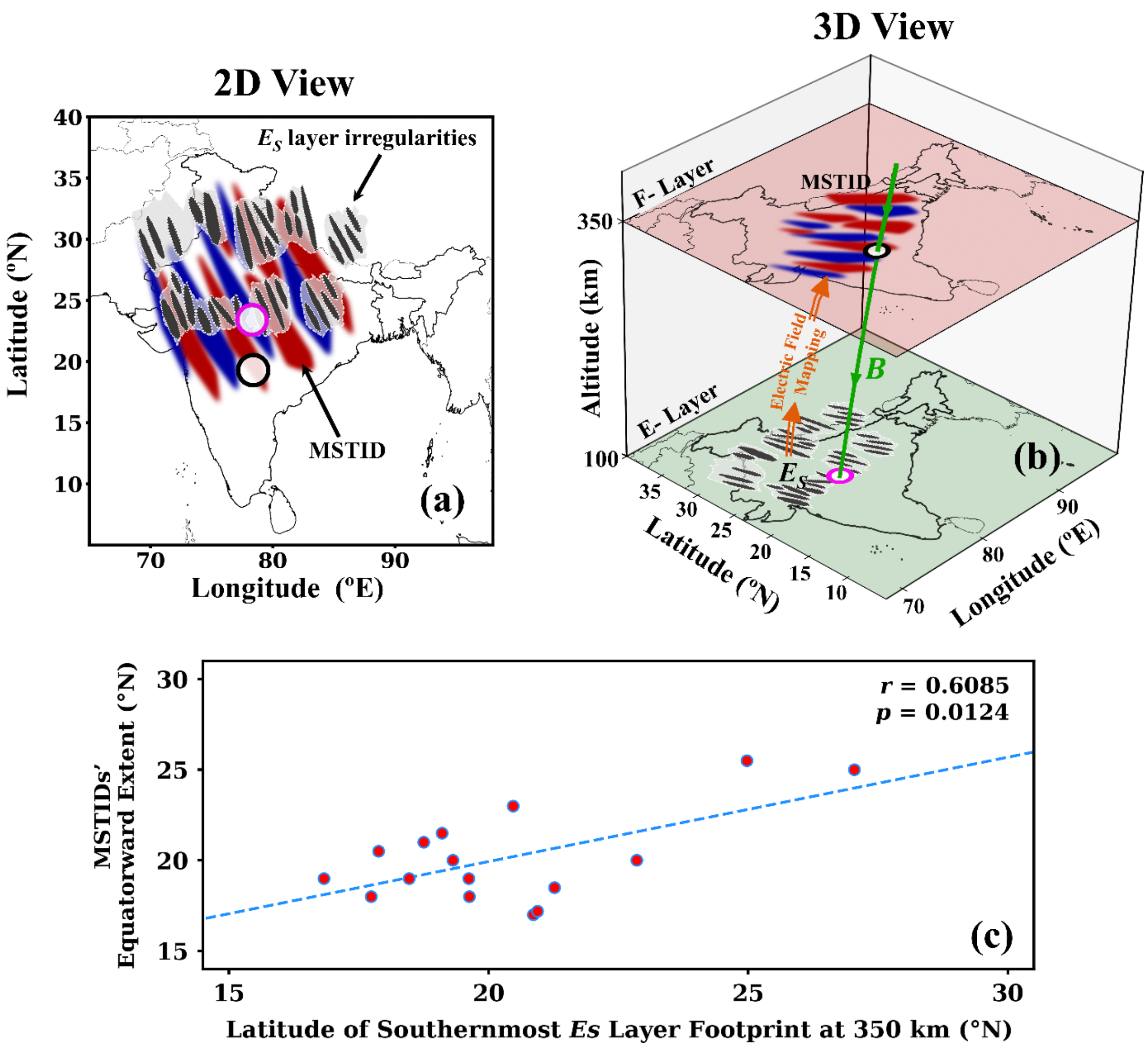

**Figure 5.** (a) Schematic illustration of F-region MSTID wavefronts (alternating maroon and blue fronts) and underlying irregular *Es*-layers (black strips) over the Indian subcontinent. (b) Three-dimensional representation showing the mapping of polarization electric fields from *Es*-layer irregularities to the F-region along geomagnetic field lines (green). White-filled circles with magenta and black boundaries denote the E-region RO-SNR tangent points and their mapped F-region footprints, respectively. (c) Scatter plot of the MSTIDs' equatorward extent versus the southernmost mapped latitude of nearby detected active *Es*-layers across different events.

The strong dependence of MSTIDs' equatorward extent on *Es*-activity through E-F coupling can also be attributed to the high solar activity during May-July 2025 (mean F10.7 ~144 sfu). Nighttime MSTID occurrence generally decreases with increasing solar activity (Candido et al., 2008; Martinis et al.,2010; Otsuka et al., 2021) because the increased neutral density enhances $\langle \nu_{in} \rangle$. Additionally, during high solar activity, higher F-region Pedersen conductivity ($\Sigma_P^F$) relative to the E-region significantly reduces $\boldsymbol{E}^*$ (Fejer et al., 1991; Otsuka et al., 2021). Both effects reduce the standalone Perkins growth rate, indicating that the MSTIDs observed in our study relied primarily on polarization electric fields mapped from *Es*-layers (Maheswaran et al., 2024). Crucially, our May-July study period coincided with the summer maximum of low and mid-latitude nighttime *Es* occurrence (Arras et al., 2008; Haldoupis et al., 2007; Lee et al., 2021). Consequently, under high solar activity and post-sunset EIA decay, the latitudinal extent of *Es*-layers acts as the primary factor governing how far nighttime MSTIDs can propagate into the low latitudes.

## 5. Conclusions

We present the first observational evidence that E-F region coupling plays an important role in regulating the equatorward ingression of nighttime MSTIDs over the Indian subcontinent during a period of high solar activity (mean F10.7~144 sfu). The study utilizes TEC measurements from a dense dual-frequency GNSS receiver network along with multi-mission RO SNR profiles and ionograms from New Delhi across 21 events. The observations show no correlation between background VTEC and the lowest latitude reached by the MSTIDs. This suggests that damping associated with background F-region plasma density near the EIA crest alone cannot explain the observed propagation limits. Instead, the equatorward propagation of MSTIDs terminates near the southernmost mapped location of the underlying *Es*-layers identified from RO-SNR profiles and supported by the ionosonde measurements. These results suggest that *Es*-mediated E-F region electrodynamic coupling may help sustain MSTID propagation and facilitates their ingression into lower latitudes, even when the uncoupled Perkins instability is suppressed during high solar activity.

## Acknowledgements

The authors acknowledge the Survey of India, Government of India, for providing the Continuously Operating Reference Stations (CORS) GNSS data. We are grateful to Gopi K. Seemala, Indian Institute of Geomagnetism, India for his support in configuring the GPS-TEC processing software. J. Upadhyaya and M. H. Jayeshkumar acknowledges the fellowship from the University Grants Commission (UGC) and Ministry of Education, Government of India respectively. R. Rathi acknowledges the fellowship from NERC grant NE/W003090/1, School of Physics and Astronomy, Lancaster University, UK. D. Patgiri's research was partly supported by an appointment to the NASA Postdoctoral Program at the NASA Goddard Space Flight Center, administered by Oak Ridge Associated Universities under contract with NASA. Arun K. Upadhayaya acknowledges the support provided by CSIR–National Physical Laboratory (CSIR-NPL) through its Unit and Laboratory In-house Project (ULIP) No. MLP002637 for carrying out this work. The work is supported by the Ministry of Education, Government of India.

**Conflict of Interest**

The authors declare no conflicts of interest relevant to this study.

**Data Availability Statement**

The Survey of India, Continuously Operating Reference Stations (SOI-CORS) data is obtained with due permission from the ftp server of CORS (ftp-cors.surveyofindia.gov.in). Additional information about the SOI-CORS network is publicly accessible through the website at https://cors.surveyofindia.gov.in. The GPS-TEC processing software utilized for this study, developed by G. Seemala, is freely available at https://seemala.blogspot.com/. The RO data from the COSMIC-2, SPIRE, MetOp-B, and MetOp-C missions were accessed through the UCAR COSMIC Data Analysis and Archive Center (CDAAC) at https://data.cosmic.ucar.edu/gnss-ro/. The Altitude-Adjusted Corrected Geomagnetic version 2 (AACGM-v2) coordinate mapping was performed using the aacgmv2 Python package, which is accessible via the Python Package Index at https://pypi.org/project/aacgmv2/. F10.7 index data are downloaded from https://omniweb.gsfc.nasa.gov/form/dx1.html. The processed gridded VTEC and DVTEC data, RO-SNR and $S_4$ profiles used in this study are publicly available on Zenodo (Upadhyaya et al., 2026).